\documentclass[12pt]{amsart}
\usepackage[english]{babel}
\usepackage{amsmath}
\usepackage{amssymb}
\usepackage{amsthm}
\usepackage{url}
\usepackage{enumitem}
\usepackage{graphicx}
\usepackage{algorithmicx,algpseudocode}
\usepackage[section]{algorithm}
\usepackage{caption}
\usepackage{wrapfig}
\usepackage{subcaption}

\theoremstyle{remark}

\usepackage{csquotes}

\DeclareMathSymbol{\minus} {\mathord}{operators}{"2D}

\usepackage[colorlinks=true, allcolors=blue]{hyperref}

\title{The CCI30 Index}
\author{Igor Rivin}
\address{Mathematics Department, Temple University and Accern Inc and The Cryptos Fund}
\email{rivin@temple.edu}
\author{Carlo Scevola}
\address{The Cryptos Fund}
\email{c.scevola@thecrytposfund.com}
\begin{document}
\begin{abstract}
We describe the design of the CCI30 cryptocurrency index.
\end{abstract}
\maketitle

\section{Introduction}

In early 2017 we decided to design an \emph{index} for the burgeoning cryptocurrency market.The purpose of this index would be twofold:
\begin{enumerate}[label=\Alph*)]
\item It would be a kind of a barometer for the cryptocurrency market.
\item It would be an investment vehicle which would allow a passive investor to participate in the growth potential of the cryptocurrency space.
\end{enumerate}
These goals meant that the index needed to be:
\begin{enumerate}[label=\Alph*)]
\item \label{diver} Diversified.
\item \label{deep} Have deep coverage of the cryptocurrency space.
\item Have as good a risk-adjusted performance profile as possible (given the points \ref{diver}, \ref{deep} above.
\end{enumerate}
Our task was complicated by a number of idiosyncrasies of the cryptocurrency market.
\begin{enumerate}[label=\arabic*]
\item The cryptocurrency space is extremely volatile (see \cite{rivin2018fear} for a more in-depth look).
\item New ``coins'' appear almost daily.
\item The lion's share of the entire cryptocurrency market capitalization\footnote{at the time of this writing, the total market capitalization of the crypto-currency market is \$302BN, of which Bitcoin and Ethereum together are worth \$180BN} is in  Bitcoin (ticker BTC) and Ethereum (ticker ETH). This would mean that a market cap weighted index (along the lines of the Standard and Poor 500 equity index) would not work very well.
\end{enumerate}
\section{solution}
\subsection{computing the index}
Between reweightings and rebalancing, the index is computed using as follows:
Suppose that there are $N$ cryptocurrencies in the index. Then, if the weights of the cryptocurrencies at (reweighting) time $T_0$ are $w_1, \dotsc, w_N,$ the value of the index at time $t$ is given by:
\[
I(t) = \sum_{i=1}^N w_i \frac{P(t)}{P(T_0)},
\]
where $P(t)$ is the price of coin $i$ at time $t.$
\subsection{Number of cryptocurrencies}
We decided to use top 30 cryptocurrencies by market capitalization - the currencies below this cutoff are not very liquid (and so their inclusion  would impair the performance of the index as an investment vehicle), and the top 30 capture 90\% of the total cryptocurrency market capitalization.
\subsection{Weighting scheme} The constituents of the index are weighted proportionally to the \emph{square root} of their market capitalizations. The square root function was chosen as a kind of a compromise - as discussed above, a market capitalization weighted index would be dominated by the top couple of cryptocurrencies, while a more slowly decaying weighting (the extreme version of which would be equal weighting) would give too much weight to the tiny (and not very liquid) crypto-currencies at the bottom of the range.
\subsection{Market capitalization computation} Market capitalization is \textbf{NOT} computed as some instantaneous number - the volatility in the cryptocurrency market is such that this would destabilize the index composition too much. Instead we use an exponentially weighted moving average of the market capitalization. In other words:
\[
M^*(T) = \dfrac{\sum{i=0}^\infty M(T-i) \exp(-\alpha i)}{\sum_{i=0}^\infty \exp(-\alpha i)},
\]
where $M(t)$ is the actual market cap at time $t,$ $M^*$ is our adjusted market cap, and $\alpha$ is a reasonable decay par2500\% success ameter. Notice that the sum goes from $0$ to $\infty$ for notational convenience - any reasonable value of $\alpha$ will make the contribution corresponding to large values of $i$ infinitesimally small.
\subsection{Rebalancing frequency} The index is completely rebalanced every quarter (on quarter boundaries), and \emph{reweighed} every month. This is necessitated by the rapid evolution of the cryptocurrency market.
\section{Performance} The performance of the CCI30 index we have designed can be seen at \url{http://cci30.com} From the beginning of 2015, the index is up by around 7000\%, with Sharpe ratio of $0.88$ (note that we compute the Sharpe Ratio using the correct formula, as explained in \cite{rivin2018sharpe}. In the same period, Bitcoin, by contrast, is up some 2500\% with a Sharpe ratio of $0.81$ (using the same formula).

\bibliographystyle{alpha}
\bibliography{sample}

\end{document}